\documentclass{article}

\usepackage{lineno,hyperref}
\modulolinenumbers[5]

\usepackage{amssymb,latexsym,amsmath,euscript}
\usepackage{amsfonts} \usepackage[cp1251]{inputenc}
\usepackage{epsfig,amssymb,amsthm} 

\usepackage{authblk}

\usepackage{enumerate}
\usepackage[english]{babel}
\newtheorem{thm}{Theorem} 
\newtheorem{lm}{Lemma}\newtheorem{cor}{Corollary}

\begin{document}

\title{On heteroclinic separators of magnetic fields in electrically conducting fluids}

\author[1]{V.~Grines}
\author[1]{T.~Medvedev}

\author[2]{O.~Pochinka}
\author[2]{E.~Zhuzhoma}

\affil[1]{Lobachevsky State University of Nizhni Novgorod, 23 Prospekt Gagarina, 603950,  Nizhni Novgorod, Russia}

\affil[2]{National Research University Higher School of Economics, 25/12 Bolshaya Pecherskaya,  603005, Nizhni Novgorod, Russia}
\maketitle
\providecommand{\keywords}[1]{\textbf{\textit{Keywords---}} #1}

\begin{abstract}
In this paper we partly solve the problem of existence of separators of a magnetic field in plasma. We single out
in plasma a 3-body with a boundary in which the movement of plasma is of
special kind which we call an (a-d)-motion. We prove that if the body is the
3-annulus or the ``fat'' orientable surface with two holes then the magnetic field
necessarily has a heteroclinic separator. The statement of the problem and the
suggested method for its solution lead to some theoretical problems from Dynamical
Systems Theory which are of interest of their own.

\end{abstract}


\section{Introduction}

The paper is about the topological structure of magnetic fields in electrically conducting fluids (collision-dominated plasmas such as stellar interiors). Such magnetic fields are common in the Universe and the problem of their origin and evolution is important for the understanding of many astrophysical and geophysical processes. It is probably safe to say that a magnetic field is a normal accompaniment of any cosmic body that is
   both fluid (wholly or in part) and rotating.
The interactions between a magnetic field and a given conducting fluid are adequately described by the equations of magnetohydrodynamics (MHD). The velocities considered are orders of magnitude smaller than the speed of light $c$.
The evolution of a magnetic field $\boldsymbol{\vec H}$ in a moving conductor is described by the induction equation
 $$ \frac{\partial\boldsymbol{\vec H}}{\partial t}=rot~\left[\boldsymbol{\vec v}\boldsymbol{\vec H}\right]+\eta\nabla^2\boldsymbol{\vec H} $$
where $\boldsymbol{\vec v}$ is the velocity of the conducting fluid and $\eta$ is the magnetic diffusivity of the medium (for more details see \cite{Alfven-book-1950-52,cowling1957magnetohydrodynamics,landau1984course}).
Literature on MHD often uses the magnetic induction $\boldsymbol{\vec B}$, for which  $\boldsymbol{\vec B}=\mu\boldsymbol{\vec H}$ where $\mu$ is the magnetic permeability of the free space.

\begin{figure}[ht]
\begin{center}\includegraphics[width=11cm]{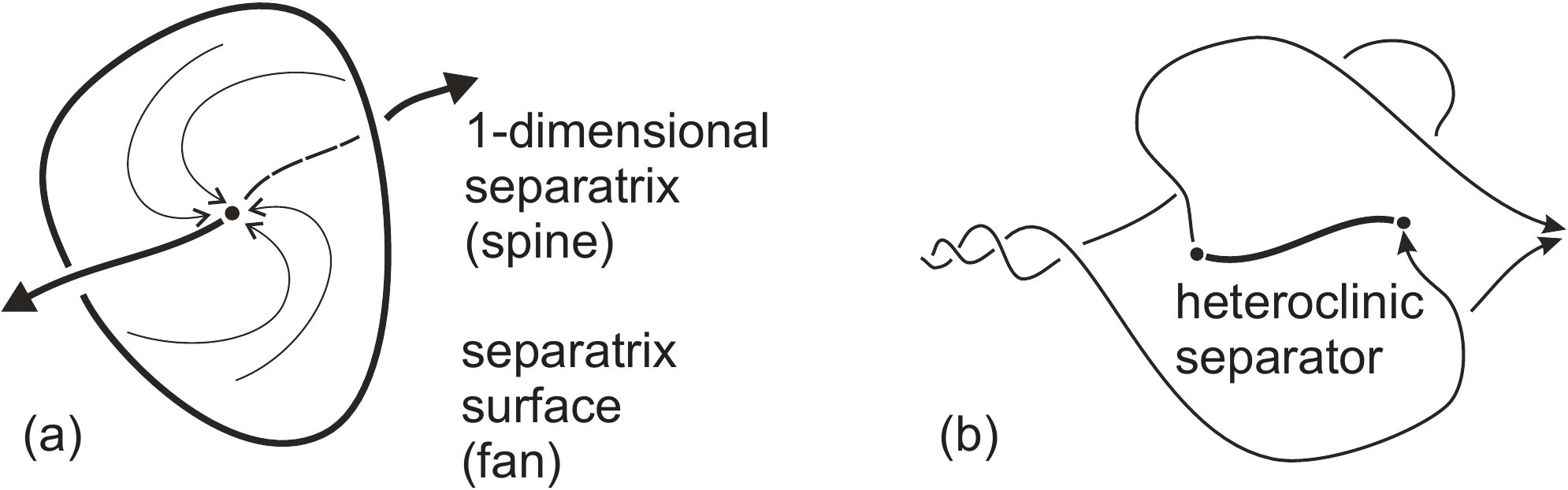}\end{center}
\caption{The structure of null point (a), and heteroclinic separator (b).}\label{spine-fan}
\end{figure}

Hannes Alfv\'{e}n \cite{alfven1943sunspots} showed that in a fluid with a large magnetic Reynolds number $\frac{1}{\eta}$ the field lines move as though they are ''frozen'' into the medium. As a consequence the topological structure of the magnetic field preserves under a short-time steady motion, though it can bifurcate under a long-time and  turbulent, in sense, motion of the medium.
The concept of frozenness implies the existence of locations where the magnetic field changes in either directions or magnitude or both
(so called $x$-points). According to {\it Amp\'{e}re's law}
  $\mu\boldsymbol{\vec j}=\boldsymbol{\nabla}\times\boldsymbol{\vec B}$, the density of the electric current $\boldsymbol{\vec j}$ is high in the neighborhood of the $x$-points and this gives rise to special configurations of the current  such as current lines and current sheets. These current configurations may include points in which the magnetic field vanishes (null points or neutral points).
Null points can occur if the magnetized medium is either a conducting plasma or a neutral gas. In plasmas, null points typically give rise to current sheets \cite{priest1982solar,priestmagnetic}. Linear null points in three dimensions typically look like a saddle of the vector field $\boldsymbol{\vec H}$. Moreover, such a saddle is a conservative one with nonzero eigenvalues $\lambda_1$, $\lambda_2$, $\lambda_3$ that satisfy the condition $\lambda_1+\lambda_2+\lambda_3=0$, due to $\nabla\cdot\boldsymbol{\vec H}=0$. Two quite distinct families of field lines pass through a null point. The null \textit{spine} is the isolated field line which approaches or recedes from the null. Its neighboring field lines form two bundles which spread out as they move away from the null and approach a surface which constitutes the null \textit{fan}, see Fig.~\ref{spine-fan} (a).  Thus, from the point of view of  Dynamical Systems Theory, a spine and fan are one-dimensional and two-dimensional separatrices respectively and the term ``separatrix'' means either of them. A magnetic line joining two nulls and representing the intersection of two fans is called a \textit{separator} \cite{priest1982solar,priestmagnetic}. A separator is \textit{heteroclinic} if it is a transversal intersection of the fans, see Fig.~\ref{spine-fan} (b).

The topological structure of a magnetic field is defined by null points, spines, fans and separators, the union of those forming the so-called \textit{skeleton} of the magnetic field. To study the global magnetic topology we have to consider first the problem of the existence of null points and separators. Experiments and observations show that the evolution of the structure of the magnetic field is similar to relaxation processes. At first plasma evolves slowly for some considerable time but at some point there occurs a topological restructuring of the magnetic configuration (reconnection) \cite{priest1982solar,priestmagnetic}. We consider the problem of the global magnetic topology under steady fluids when the skeleton is invariant (before reconnection).

We suggest to approach this problem in the following way. We consider  a body in plasma (i.e. a part of plasma) of special kind and we study movements of plasma such that all the boundary components of the body move inside or outside of the body so that after some time interval all the boundary components are parallel to the initial boundary components (see exact definitions below).  The statement of the problem implies that during the movement the topological structure of the magnetic field remains unchanged, therefore it is natural to assume the skeleton of the magnetic field to be invariant with respect to the movement of plasma. Notice that we do not demand all the points of the skeleton to be fixed, we just assume that the points on the skeleton, while moving, remain on the skeleton. We also assume (and this is the only essential restriction) that the null points are not only null hyperbolic points of the magnetic field but they also are null hyperbolic points of the movement of plasma. Taking into consideration that the Kupka-Smale theorem from Dynamical System Theory states that for every typical movement all periodical points  (including fixed points) are hyperbolic \cite{Ni}, we consider only the class of typical movements of plasma.
The suggested approach lets us apply methods and results of Dynamical Systems Theory because it turns out to be possible to extend the motion of plasma to some 3-manifold in such a way that one gets a classical dynamical system. Though doing so  we loose information on the structure of the magnetic field outside the considered body,  on the other hand we get an instrument to study its structure in some part of space. Besides, the global structure on the considered manifold sheds light on possible real structures of the magnetic field. The suggested approach to the problem and the method lead to theoretical issues of Dynamical Systems Theory whose solution is of interest in its own.

\section{Formulation of the main results}

Let $M^2_p$ be an orientable closed surface of genus $p\geq 0$ smoothly embedded in $\mathbb{R}^3$. Two surfaces $M^2_{p_1,1}$, $M^2_{p_2,2}\subset\mathbb{R}^3$ are called \textit{parallel} if they bound the set which is homeomorphic to $M^2_p\times (0;1)$. As a consequence, $p_1=p_2=p$ and $M^2_{p,1}\cap M^2_{p,2}=\emptyset$. The closure of $M^2_p\times [0;1]$, in short
a \textit{fat surface}, is a 3-manifold with the boundary $M^2_{p,1}\cup M^2_{p,2}$. In particular, $\mathcal{S}={\mathbb S}^2\times [0;1]$ is a \textit{fat sphere} where ${\mathbb S}^2$ is the 2-sphere.
One of its boundary 2-spheres, say $S^2\times\{0\}=S_{int}$, bounds an open 3-ball in $\mathbb{R}^3$ which is disjoint from $\mathcal{S}$. We call $S_{int}$ \textit{interior} while the sphere $S^2\times\{1\}=S_{ext}$ is called \textit{exterior}. Let $B_{int}$ and $B_{ext}$ be the 3-balls bounded by $S_{int}$ and $S_{ext}$ respectively.

	Let $G_p$ be a fat surface bounded by parallel surfaces  $M^2_{p,1}$ and $M^2_{p,2}$ of genus $p\geq 1$ and let it contain two 2-spheres $S_1$ and $S_2$ such that $S_1\cap S_2=\emptyset$ and the balls $B_1$ and $B_2$, bounded by $S_1$ and $S_2$ respectively, do not intersect. Denote by $\mathcal{M}$ the fat surface $G_p$ with two holes, i.e. the set $G_p\setminus\{ B_1\cup B_2\}$.

Let $M$ be either $\mathcal{S}$ or $\mathcal{M}$ which is smoothly embedded into $\mathbb{R}^3$. Suppose $M$ be a part of space with plasma of some astrophysical object with a magnetic field  $\boldsymbol{\vec B}$. Denote by $\boldsymbol{\vec B}_0$ the restriction of $\boldsymbol{\vec B}$ to $M$ and let all the null points of $\boldsymbol{\vec B}$ be typical (i.e. hyperbolic). As a consequence we have that $M$ has only finitely many null points.

We assume that
\begin{enumerate}
\item the separatrices of the null points intersect transversally (if at all);
\item  if $M$ is the fat surface with holes $\mathcal{M}$ the separatrices intersect the components $S_1$ and $S_2$ transversally (if intersect at all).
\end{enumerate}

\begin{figure}[ht]
\begin{center}\includegraphics[width=11cm]{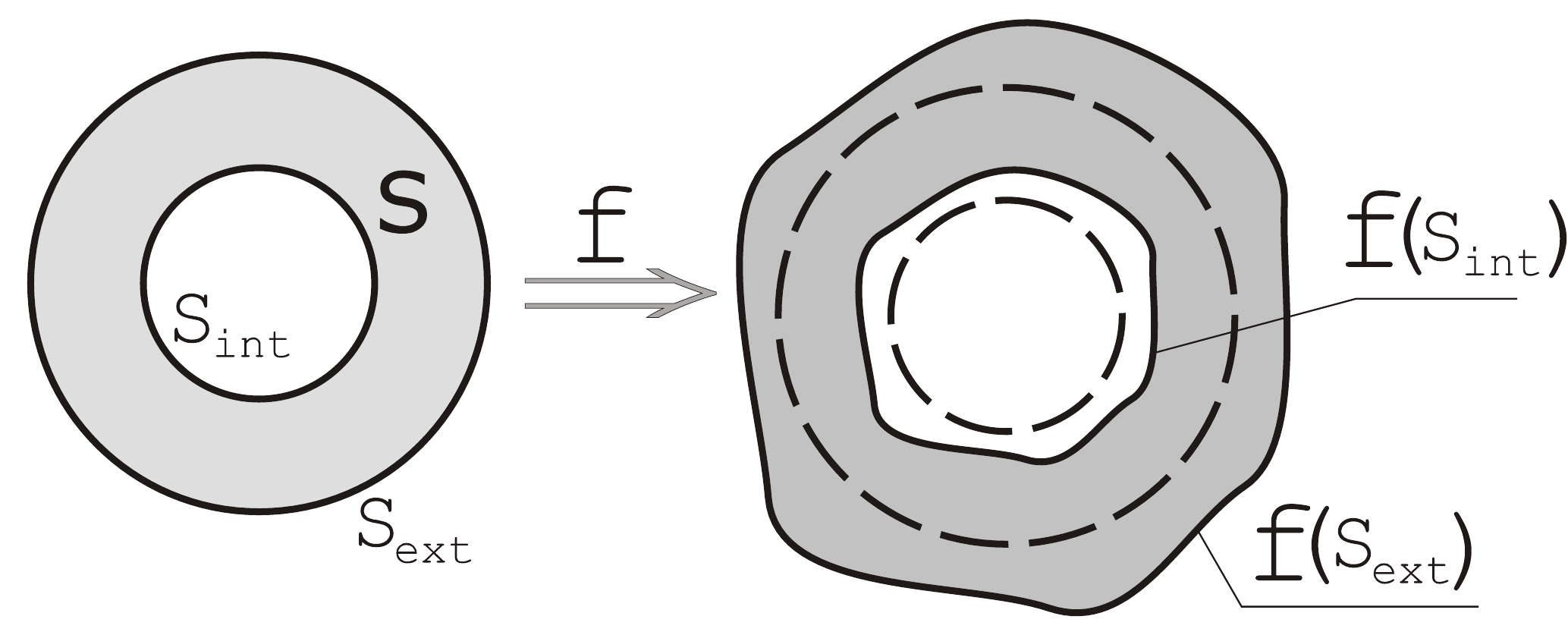}\end{center}
\caption{ (\ref{a}-\ref{d})-motion of the fat sphere $\mathcal{S}$.}\label{moving}
\end{figure}

A map
 $ f_0: M\to f_0(M)\subset\mathbb{R}^3 $
is said to be an (\ref{a}-\ref{d})-\textit{motion} if it satisfies the following conditions:

\begin{enumerate}[(a)]
  \item$f_0$ is an orientation preserving diffeomorphism to its image. The non-wandering set of $f_0$ consists of fixed hyperbolic points which coincide with the null points of the vector field  $\boldsymbol{\vec B}_0$.\label{a}
  \item The boundary components of $f_0(M)$ are pairwise disjoint from the boundary components of  $M$.\label{b}
  \item If $M$ is the fat sphere $\mathcal{S}$ then the interior boundary sphere maps inside $\mathcal{S}$ while the external boundary sphere maps outside $\mathcal{S}$, i.e. $f_0(S_{int})\subset\mathcal{S}$ and $f_0(S_{ext})\subset\mathbb{R}^3\setminus\left(\mathcal{S}\cup B_{int}\right)$  (see Figure~\ref{moving}). 

If $M$ is the fat surface with holes $\mathcal{M}$ then one of the boundary spheres, say $S_1$, maps inside $\mathcal{M}$ while the other boundary sphere $S_2$ maps outside $\mathcal{M}$,  one of the boundary surfaces $M^2_{p,i}$, say $M^2_{p,1}$, maps inside $\mathcal{M}$ while the other boundary surface $M^2_{p,2}$ maps outside $\mathcal{M}$ and the restriction $f_0|_{M^2_{p,i}}: M^2_{p,i}\to f_0(M^2_{p,i})$ is  homotopy trivial for each $i=1,2$. Moreover, every $x\in M^2_{p,1}$ which does not belong to a separatrix leaves $\mathcal{M}$ through $S_2$, i.e. $f_0^k(x)\in B_2$ for some $k>1$ whereas for the reverse motion every $x\in M^2_{p,2}$ which does not belong to a separatrix leaves $\mathcal{M}$ through $S_1$, i.e. $f_0^{-k}(x)\in B_1$ for some $k>1$.\label{c}
  \item The fans and the spines are invariant with respect to $f_0$ and the fixed points of  $f_0$ are of the same type as the null points of the field
           $\boldsymbol{\vec B}_0$.\label{d}
\end{enumerate}

Some explanation on homotopy triviality. Unlike sphere for which there is only one homotopy class of orientation preserving diffeomorphisms, surfaces of genus greater than zero have a countable family of such classes. Since the surfaces $M^2_{p,i},\ i=1,2$ are parallel, the generators of their fundamental groups are naturally isomorphic. The homotopy triviality then means that the restrictions $f_0|_{M^2_{p,i}}$ are homotopy identical.

Notice that we do not assume that the magnetic lines of the field $\boldsymbol{\vec B}_0$ intersect the boundary components of $M$ transversally. Therefore, the fans and the spines can, in general, intersect the boundary components of $M$ by several connected components. This mathematical model physically means that we consider a motion of plasma in a period  during which the singular points and their fans and spins are preserved. If follows from these properties that the separators (if they exist) are invariant with respect to $f_0$ and their number (including zero) does not change during this time period.

Our main results for an (\ref{a}-\ref{d})-motion of a fat sphere of plasma are the following.

\begin{thm}\label{thm:singularities-at-least-two-exist-separator}
Let $f_0: \mathcal{S}\to f_0(\mathcal{S})\subset\mathbb{R}^3$ be an (\ref{a}-\ref{d})-motion of a fat sphere $\mathcal{S}$ of plasma with a magnetic field $\boldsymbol{\vec B}_0$.
Let $\boldsymbol{\vec B}_0$ have in $\mathcal{S}$ null points. Then the number of these points is even (and there are at least two of them), their fans intersect and there are finitely many (not zero) heteroclinic separators.
\end{thm}

\begin{cor}\label{cor:from-thm-singularities-at-least-two-exist-separator}
Let the assumptions of Theorem \ref{thm:singularities-at-least-two-exist-separator} hold and assume that the spines of the magnetic field $\boldsymbol{\vec B}_0$ do not intersect the fans in $\mathcal{S}$. Then the fan of each null point contains at least one heteroclinic separator in $\mathcal{S}$.
\end{cor}

For an (\ref{a}-\ref{d})-motion of a fat surface $\mathcal{M}$ of plasma the results are the following.

\begin{thm}\label{thm:fat-torus-exist-separator-singularities}
Let $f_0: \mathcal{M}\to f_0(\mathcal{M})\subset\mathbb{R}^3$ be an (\ref{a}-\ref{d})-motion of a fat surface  $\mathcal{M}$ which belongs to some domain of plasma with a magnetic field
$\boldsymbol{\vec B}_0$. Then the field $\boldsymbol{\vec B}_0$ has in $\mathcal{M}$ at least two null points such that their fans intersect and there are finitely many (more than zero) heteroclinic separators.
\end{thm}

\begin{cor}\label{cor:from-thmfat-torus-exist-separator-singularities}
Let the assumptions of Theorem \ref{thm:fat-torus-exist-separator-singularities} hold and assume that the spines  of the null points of the vector field $\boldsymbol{\vec B}_0$ do not intersect the fans  in $\mathcal{M}$. The fan of each null point has in $\mathcal{M}$ at least one heteroclinic separator.
\end{cor}

\section{Proof of the main results}

We now recall some notions and facts about Morse-Smale diffeomorphisms. More detailed explanations can be found  for example in \cite{Ni,Grines-Pochinka-book-2011,anosov1988encyclopaedia, S3}. Denote the set of the non-wandering points of a diffeomorphism $f$ by $NW(f)$.
For $p\in NW(f)$ denote by $W^s(p)$ (by $W^u(p)$) the stable (respectively unstable) manifold of the point $p$. A diffeomorphism $f$ is said to be \textit{Morse-Smale} if its non-wandering set $NW(f)$ is hyperbolic,  $NW(f)$ consists of finitely many points and for any two distinct points $p$, $q\in NW(f)$ the invariant manifolds $W^s(p)$, $W^u(q)$ intersect transversally (if intersect at all). A Morse-Smale diffeomorphism is said to be \textit{gradient-like}, if for any two periodic points  $p$, $q\in NW(f)$ from $W^u(p)\cap W^s(q)\neq \emptyset$ it follows that $\dim W^s(p) < \dim W^s(q)$.
A point $x\in M$ is said to be \textit{heteroclinic} if $x$ is the point of transversal intersection of invariant manifolds $W^s(p)$, $W^u(q)$ where  $p$, $q\in NW(f)$ and $\dim W^s(p) = \dim W^s(q)$. It is proved that a Morse-Smale diffeomorphism is gradient-like if and only if it has no heteroclinic points \cite{S3}. A Morse-Smale diffeomorphism is said to be  \textit{polar} if its non-wandering set consists of exactly one source and one sink.

If $W^u(p)\cap W^s(q)\neq\emptyset$ and $dim~W^s(p) < dim~W^s(q)$ then a connected component of $W^u(p)\cap W^s(q)$ is said to be
\textit{heteroclinic submanifold}. If the dimension of the manifold equals  $3$ then every heteroclinic submanifold is either a simple closed curve (which is homeomorphic to the circle) or a non-closed curve without self-intersections (homeomorphic to the open interval). We call such curves \textit{heteroclinic}.

Let $p$ be a periodic point of an orientation preserving Morse-Smale diffeomorphism  $f$. The \textit{Morse index} $u(p)$ of $p$ is the topological dimension of the unstable manifold  $W^u(p)$, $u(p)\stackrel{\rm def}{=}\dim W^u(p)$. The \textit{Kronecker-Poincar\'{e} index} of $p$ is
$ind~(p,f)\stackrel{def}{=}(-1)^{u(p)}$.

Denote by $S^3$ the 3-sphere. The following lemma is of key importance for the proof of Theorem \ref{thm:singularities-at-least-two-exist-separator}.
\begin{lm}\label{lm:f-is-morse-smale}
There is an embedding $\mathcal{S}\subset S^3$ and there is an extension of $f_0$ to a polar Morse-Smale diffeomorphism $f: S^3\to S^3$ such that the non-wandering set $NW(f)$ is the union of a source, a sink and the fixed points of $f_0$.
\end{lm}

\textsl{Proof}. 
Glue the balls $B_{int}$, $B_{ext}$ to the respective boundary components $S_{int}$, $S_{ext}$ of the fat sphere $\mathcal{S}$. Then we get the closed manifold $S^3=\mathcal{S}\cup B_{int}\cup B_{ext}$ diffeomorphic to the 3-sphere and we get the natural embedding $\mathcal{S}\subset S^3$. From the properties of the diffeomorphism
$f_0: \mathcal{S}\to f_0(\mathcal{S})$ it follows that the 2-sphere $S_{int}$ maps inside the fat sphere $\mathcal{S}$. Therefore $f_0$ can be extended to $B_{int}$ in such a way that a hyperbolic source appears in the interior of $B_{int}$. Analogously $f_0$ can be extended to the ball $B_{ext}$ so that a hyperbolic sink appears in the interior of $B_{ext}$. Denote this extension of $f_0$ by $f: S^3\to S^3$. Clearly $f_0$ can be extended in such a way that $f$ would be a diffeomorphism whose non-wandering set is the union of the non-wandering set of the diffeomorphism $f_0$ and the two new fixed points.

Thus, the non-wandering set of the diffeomorphism $f$ is finite and it consists of hyperbolic fixed points. Since the separatrices of the fixed points intersect transversally by the assumption, the diffeomorphism $f$ is Morse-Smale. Moreover, $f$ is polar because it has only two node fixed points.$\Box$

The following lemma is the base for the proof of Theorem \ref{thm:fat-torus-exist-separator-singularities}. Recall that $S^1$ denotes the circle and  $M^2_p$ denotes the closed orientable surface of genus $p$. Everywhere below we assume $p\geq 1$.
\begin{lm}\label{lm:extention-to-three-torus}
There is an embedding $\mathcal{M}\subset M^2_p\times S^1$ and there is an extension of $f_0$ to a polar Morse-Smale diffeomorphism $f: M^2_p\times S^1\to M^2_p\times S^1$ such that the non-wandering set $NW(f)$ is the union of a source, a sink and the non-wandering points of the diffeomorphism  $f_0$.
\end{lm}

\textsl{Proof}. 
Glue balls $B_1$, $B_2$ to the respective boundary components $S_1$, $S_2$ of the fat surface $\mathcal{M}$. Then we get the body
$\mathcal{M}\cup B_1\cup B_2$ and the natural embedding $\mathcal{M}\subset\mathcal{M}\cup B_1\cup B_2$. By the condition (\ref{d}) the 2-sphere $S_1$ maps inside $\mathcal{M}$. Therefore $f_0$ can be extended to the ball $B_1$ in such a way that a hyperbolic source appears inside $B_1$. Analogously $f_0$ can be extended to the ball $B_2$ so that a hyperbolic sink appears inside $B_2$. The body $\mathcal{M}\cup B_1\cup B_2$ is homeomorphic to the direct product  $M^2_p\times [0;1]$ with the boundary components $T_1=M^2_p\times\{0\}$, $T_2=M^2_p\times\{1\}$, therefore below we identify $\mathcal{M}\cup B_1\cup B_2$ and $M^2_p\times [0;1]$. It is known that from homotopy triviality of the restrictions $f_0|_{T_1}$, $f_0|_{T_2}$ it follows that each of these restrictions is isotopic to the identity map. Hence, there is $\varepsilon>0$ and there is an extension of $f_0$ to the body
$M^2_p\times [-\varepsilon ;1+\varepsilon ]$ such that the restrictions $f_0|_{M^2_p\times\{-\varepsilon\}}$ and  $f_0|_{M^2_p\times\{1+\varepsilon\}}$ are the identities on the first coordinate. Without loss of generality one can assume each of  $f_0|_{M^2_p\times\{-\varepsilon\}}$, $f_0|_{M^2_p\times\{1+\varepsilon\}}$ to be a shift along the second coordinate $\{\cdot\}\times\{-\varepsilon\}\to\{\cdot\}\times\{-\varepsilon+\delta\}$ and $\{\cdot\}\times\{1+\varepsilon\}\to\{\cdot\}\times\{1+\varepsilon+\delta\}$ respectively.

Having glued the boundary components $M^2_p\times\{-\varepsilon\}$, $M^2_p\times\{1+\varepsilon\}$ by the identity map we get the manifold $M^2_p\times S^1$ from
$M^2_p\times [-\varepsilon ;1+\varepsilon ]$. It follows from the previous construction that there is the desired extension of the diffeomorphism $f_0$ to $M^2_p\times S^1$.
$\Box$

The result below follows from the finiteness of every heteroclinic chain, so called ``beh'' (see also \cite{BoGr}, details of the proof we leave to the reader).
\begin{lm}\label{one-more-simple1}
Let $M^3$ be a closed orientable 3-manifold and let $f: M^3\to M^3$ be a polar Morse-Smale diffeomorphism with a nonempty set of saddle periodic points of Morse indexes 1 and 2. Let $\alpha$ be the source and  $\omega$ be the sink of $f$. Then there are saddle periodic points $\sigma_1$, $\sigma_2$ of Morse indexes $u(\sigma_1)=1$ and  $u(\sigma_2)=2$ respectively such that the following inclusions hold
 $$ W^u(\sigma_1)\setminus\{\sigma_1\}\subset W^s(\omega),\quad  W^s(\sigma_2)\setminus\{\sigma_2\} \subset W^u(\alpha ). $$
\end{lm}

\begin{lm}\label{simple2}
Let the assumptions of Lemma \ref{one-more-simple1} be satisfied. Then the sets
 $$ C_{\omega}\stackrel{\rm def}{=}\{\omega \}\cup W^u(\sigma_1) \mbox{ and } C_{\alpha}\stackrel{\rm def}{=}\{\alpha \}\cup W^s(\sigma _2) $$
are embeddings of the circles.
\end{lm}

\begin{figure}[h]
\begin{center}\includegraphics[width=9cm]{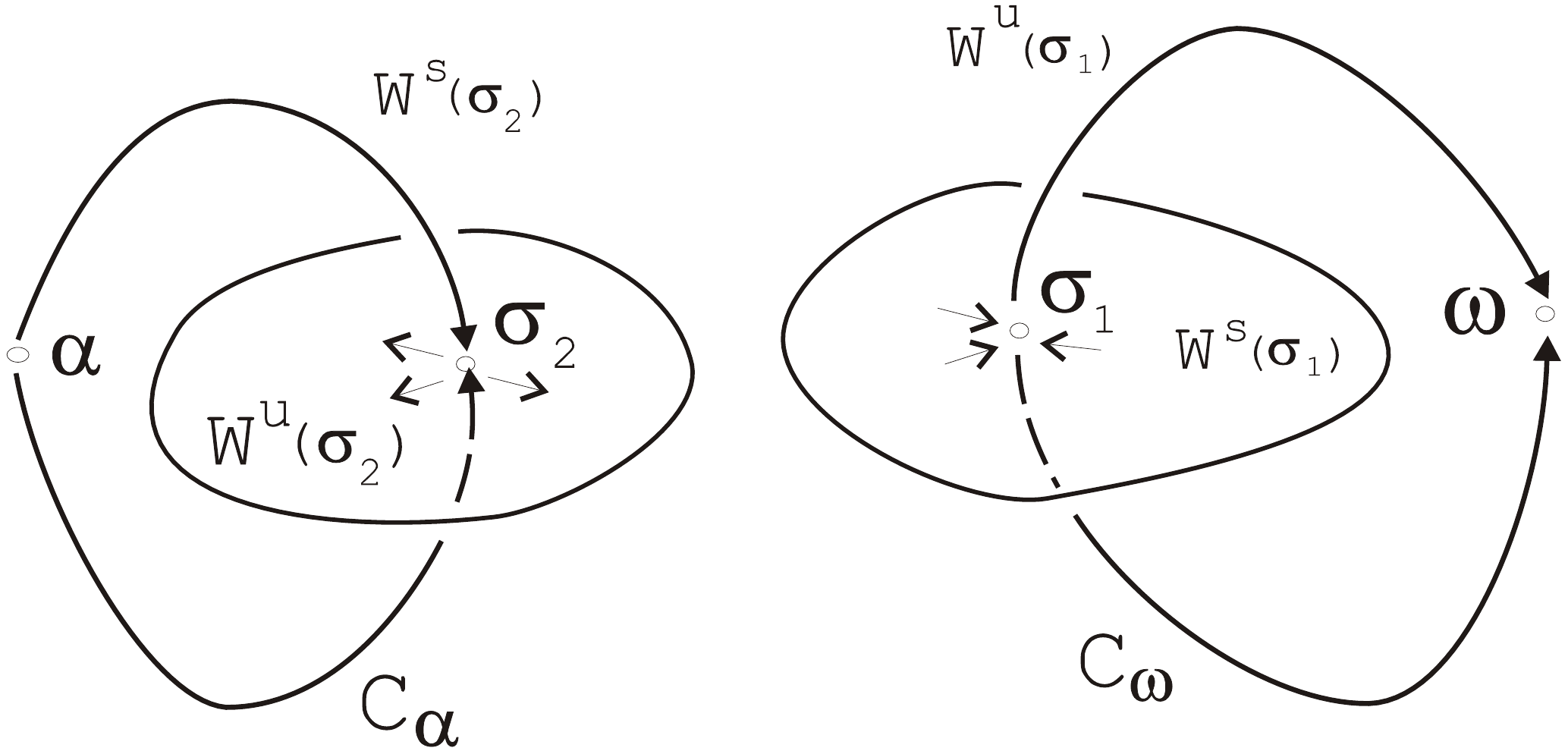}\end{center}
\caption{The curves $C_{\alpha}$ and $C_{\omega}$.}\label{fig31}
\end{figure}

\textit{Proof.}
We prove the lemma only for $C_{\omega}$ because for $C_{\alpha}$ the proof is analogous.
There is a $C^1$-immersion $\varphi : {\mathbb R}\to W^u(\sigma _1)$, where $\varphi (0) = \sigma _1$, which is a one-to-one map to its image. We now show that the immersion $\varphi$ extends to a homeomorphism
 $$ \varphi : S^1\cong {\mathbb R}\cup \{\infty \}\to W^u(\sigma _1)\cup \{\sigma _1\}, $$
if we set $\varphi (\pm \infty ) = \omega$. Since $W^u(\sigma _1) \setminus \sigma _1\subset W^s(\omega )$ the desired assertion follows from the fact that the
$\omega$-limit set of every point from the set $W^u(\sigma _1)\setminus \sigma _1$ is the point $\omega$, see figure~\ref{fig31}.
$\Box$

Let $M^3$ be a smooth orientable closed  3-manifold. A 2-sphere $S^{2}$ in $M^3$ is said to be \textit{cylindrical} or \textit{cylindrically embedded} into $M^3$, if there is a topological embedding $h: \mathbb{S}^{2} \times [-1,1] \to M^3$ such that $h(\mathbb{S}^{2} \times\{0 \}) = S^{2}$. A manifold $M^3$ is said to be \textit{irreducible}, if every 2-sphere, cylindrically embedded into $M^3$, bounds a 3-ball in $M^3$. The manifolds $S^3$ and $M^2_p\times S^1$ are known to be irreducible. Therefore, to prove the main results we have to answer the question if a polar Morse-Smale diffeomorphism on an irreducible orientable closed 3-manifold can have heteroclinic curves.

It follows from \cite{BGMP} that every Morse-Smale diffeomorphism on any irreducible 3-manifold except the sphere necessarily has heteroclinic curves (compact or noncompact). The following theorem generalizes this result for polar  diffeomorphisms.

\begin{thm} \label{0.sphere1} Let $f: M^3\to M^3$ be a polar Morse-Smale diffeomorphism on an irreducible closed orientable 3-manifold $M^3$ and let $f$ have at least one saddle point. Then there is a saddle periodic point such that its invariant 2-manifold contains non-compact heteroclinic curve.
\end{thm}

\textsl{Proof}. 
Denote the source, the sink and a saddle of the diffeomorphism $f$ by $\alpha,\omega$ and $\sigma$ respectively. Without loss of generality the point $\sigma=\sigma_1$ can be assumed to satisfy the conditions of Lemma \ref{simple2} (i.e. the manifold $W^s_\sigma$ can be assumed to be of dimension 2, otherwise one can consider a suitable degree of $f$).
We now show that $W^s_\sigma$ contains a non-compact heteroclinic curve. Assume the contrary, that either the manifold $W^s_\sigma$ contains no heteroclinic curves or all these curves are compact. Then there is a simple smooth closed curve $\gamma\subset W^s_\sigma$ which is disjoint from the heteroclinic curves and which bounds a 2-disk $d_\gamma\subset W^s_\sigma$ such that the point $\sigma$ lies in its interior. According to \cite{S3}, $M^3=\bigcup\limits_{p\in\Omega_f}W^u_p=\bigcup\limits_{p\in\Omega_f}W^s_p$ and therefore $\gamma\subset W^u_\alpha$. By assumption  $cl~W^u_\sigma=W^u_\alpha\cup\omega$ and $cl~W^u_\sigma$ is a topologically embedded circle while $W^u_\sigma$ is a smooth curve. Now making use of the disk $d_\gamma$ we are going to prove that in $M^3$ there is a cylindrically embedded 2-sphere $S$ which contains $d_\gamma$ and which transversally intersects the circle $cl~W^u_\sigma$ at the unique point $\sigma$. This would contradict the fact that the intersection $cl~W^u_\sigma\cap S$ must consist of even number of points because the sphere $S$ bounds a 3-ball in the irreducible manifold $M^3$.

{\bf Construction of the sphere $S$.} Consider in $W^u_\alpha$ a smooth  $3$-ball $\tilde B$ containing $\alpha$. Since $\gamma\subset W^u_\alpha$ there is a natural $k$ for which $f^{-k}(\gamma)\subset int~\tilde B$. Set $\tilde\gamma=f^{-k}(\gamma)$ and $Q=f^{-k}(d_\gamma)$. Pick a natural $k_0$ such that the 3-ball $B_0=f^{-k_0}(\tilde B)$ is inside the ball $\tilde B$.  Set $\gamma_0=f^{-k_0}(\tilde\gamma)$ and $\Sigma=f^{-k_0}(Q)$. Without loss of generality the 2-disk $\Sigma$ can be assumed to be transversal to the 2-spheres $\tilde S=\partial\tilde B$ and  $S_0=\partial B_0$. By construction the smooth ball $B_0$ satisfies the following conditions:

\begin{itemize}
\item[(*)] $B_0\subset\tilde B\subset W^u_\alpha$, $\alpha\in B_0$, $B_0\cap Q=\emptyset$, the 2-sphere $S_0$ is transversal to $\Sigma$.
\end{itemize}

There are two cases: 1) every curve in the set $\Sigma\cap S_0$ separates $\gamma_0$ and $\sigma$ in $\Sigma$; 2) there is a curve in  $\Sigma\cap S_0$ which does not separate $\gamma_0$ and $\sigma$ in $\Sigma$. We now construct the desired sphere $S$ in each case.

1) If every curve in the set $\Sigma\cap S_0$ separates $\gamma_0$ and $\sigma$ in $\Sigma$ then by construction the  number of such curves is odd. Denote them by $\{c_1,\dots,c_m\}$. Up to renumeration of $c_i$ one can assume that $\Sigma\setminus (c_1\cup\dots\cup c_m)$ consists of finitely many connected components $\Delta_{1},\Delta_{2},\dots,\Delta_{k+1}$ such that $\Delta _ {1} $ is homeomorphic to the 2-disk bounded by
$c_1$ and such that $\sigma\in\Delta _ {1} $; for $2\leq i \leq k$ the component $\Delta_i$ is homeomorphic to the 2-annulus bounded by $c_ {i-1} $ and $c_i$; the annulus $\Delta_{k+1}$ is bounded by the curves $c_k$ and $\gamma_0$. Let $d_1\subset S_0$ be the 2-disk bounded by the curve $c_1$. Then $S=\Delta_1\cup d_1$ is the desired sphere.

\begin{figure}[h]
\begin{center}\includegraphics[width=9cm]{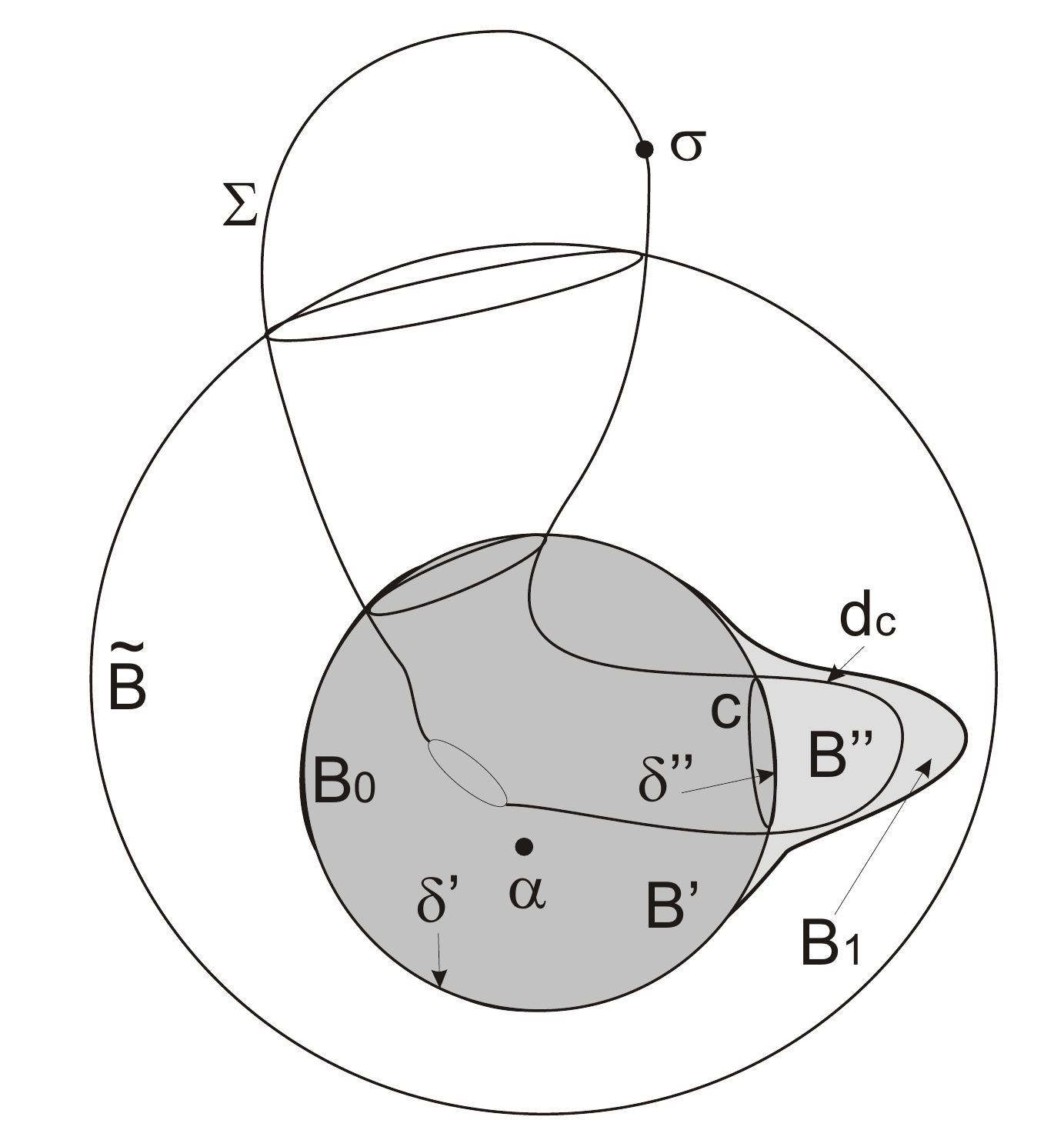}\end{center}
\caption{Illustration to the proof of Theorem \ref{0.sphere1}}\label{ad1}
\end{figure}

2) Let there be a curve in $\Sigma\cap S_0$ which does not separate $\gamma_0$ and $\sigma$ in $\Sigma$. Denote by  $\mathcal C_{B_0}$ the set of the curves in the intersection $\Sigma\cap S_0$ and denote by $\mathcal C^+_{B_0}$ ($\mathcal C^-_{B_0}$) the curves from $\mathcal C_{B_0}$ which separate (do not separate) $\gamma_0$ and $\sigma$ in $\Sigma$.  Construct a ball  $B_1$ such that it satisfies $(*)$ and such that the number of the curves in  $\mathcal C^-_{B_1}$ is  less than the number of the curves in the set $\mathcal C^-_{B_0}$.

Every curve of the set $\mathcal C^-_{B_0}$ bounds on $\Sigma$ a disk which contains neither $\gamma_0$ nor $\sigma$. Let  $c\in\mathcal C^-_{B_0}$ be the innermost curve, i.e. $c$ bounds on $\Sigma$ the disk $d_c$ that contains neither $\gamma_0$ nor $\sigma$ and besides $int~d_c$ contains no curves of the set $\mathcal C^-_{B_0}$. Since $c\cap Q=\emptyset$ and $\sigma\in Q$ we have $d_c\cap Q=\emptyset$ and therefore $d_c\subset\tilde B$. On the other hand $c$ bounds on $S_0$  2-disks $\delta'$ and $\delta''$. Notice that $S' =d_c\cup\delta'$ and $S''=d_c\cup\delta''$ are tamely embedded 2-spheres lying inside $\tilde B$. Denote by $B'$ and $B''$ the $3$-balls which are contained in $\tilde B$ and which are bounded by $S'$ and $S''$ respectively (see figure \ref{ad1}). Since $S'\cap Q=\emptyset$, $S''\cap Q=\emptyset$ and $y_0\in Q$ we have $B'\cap Q=\emptyset$, $B''\cap Q=\emptyset$. And since $(B'\cup B'')\supset B_0$ we have $\alpha\in B'$ or $\alpha\in B''$.

To be definite let $\alpha\in B'$. Then we can pick a smooth ball $B_1$ which is arbitrarily close to $B'$, which satisfies $(*)$ and such that
$\mathcal C^-_{B_1}=\mathcal C^-_{B_0}\setminus c$. We continue this process and get a ball $B$ which satisfies $(*)$ and for which $\mathcal C^-_{B}=\emptyset$.
$\Box$

\begin{cor}\label{l.sphere1} The invariant 2-manifold of every saddle point of a polar gradient-like diffeomorphism on an irreducible manifold contains a non-compact heteroclinic curve.
\end{cor}

\textsl{Proof}. 
 The assertion is immediate from the proof of Theorem \ref{0.sphere1} and the fact that every one-dimensional separatrix of a gradient-like diffeomorphism is disjoint from the 2-dimensional separatrices.
$\Box$

\textsl{Proof of Theorem \ref{thm:singularities-at-least-two-exist-separator}.}
Taking into account Lemma \ref{lm:f-is-morse-smale}, to prove that the number of null points is even it is enough to show that a Morse-Smale diffeomorphism $f: M^3\to M^3$ necessarily has even number of periodic points on the closed 3-manifold $M^3$. Since $f$ preserves orientation of the ambient manifold by condition (\ref{a}), we can assume it to preserve orientation of all separatrices of the saddle fixed points (otherwise we consider a suitable iteration of $f$). Such polar diffeomorphism of the 3-sphere has the Lefschetz number  equal to zero. From this and from the Lefschetz formula it follows that the number of the saddle fixed points of Morse index 1 equals the number of the saddle fixed points of Morse index 2. Hence, $f$ has even number of the periodic points. Notice that we have also proved that $f$ necessarily have saddle points of distinct Morse index. Now Theorem \ref{thm:singularities-at-least-two-exist-separator} follows from Theorem \ref{0.sphere1}.
$\Box$

\textsl{Proof of Theorem \ref{thm:fat-torus-exist-separator-singularities}.}
First we show that  $f$ has at least one saddle point. Suppose the contrary, then $f$ is a diffeomorphism of ``source-sink'' type. According to \cite{GrMePoZh} the ambient manifold is then homeomorphic to the 3-sphere. This contradiction proves that  $f$ has at least one saddle point.

Since $f$ preserves orientation of the ambient manifold by condition (\ref{a}), we can assume it to preserve orientation of all separatrices of the saddle fixed points (otherwise we consider a suitable iteration of $f$). According to the condition (\ref{d}) for $f_0: M^2_p\times S^1\to f_0(M^2_p\times S^1)$ the diffeomorphism $f$ is homotopy trivial (i.e. it induces the identity map of the fundamental group of the torus). Therefore, its Lefschetz number equals to zero. The sum of the Morse indexes of the source and of the sink equals to zero, $(-1)^0+(-1)^3=0$. From this and from the Lefschetz formula it follows that existence of one saddle fixed point courses existence of another fixed point with different Morse index. Now Theorem \ref{thm:fat-torus-exist-separator-singularities} follows from Theorem \ref{0.sphere1}.
$\Box$

Corollaries \ref{cor:from-thm-singularities-at-least-two-exist-separator}, \ref{cor:from-thmfat-torus-exist-separator-singularities} follow from the proofs of Theorems \ref{thm:singularities-at-least-two-exist-separator}, \ref{thm:fat-torus-exist-separator-singularities} and Corollary \ref{l.sphere1}.

\section*{Acknowledgments} This work was supported by the Russian Foundation for Basic Research, Grants 12-01-00672-a and 13-01-12452-ofi-m, and partly by the Russian Science Foundation, Grant 14-11-00446. We especially thank Konstantin Kirsenko for financial support. We also thank the unknown reviewer for his(her) useful suggestions.

\bibliography{literatura}

\end{document}